\documentclass{mn2e}
\usepackage{graphicx}
\usepackage{amsbsy}
\usepackage{amsmath}
\usepackage{epsf}
\usepackage{paralist}
\usepackage{tablefootnote}

\newcommand{\apj}{ApJ}
\newcommand{\apjl}{ApJL}
\newcommand{\mnras}{MNRAS}
\newcommand{\apjs}{ApJS}

\newcommand{\aap}{A\&A}

\newcommand{\nat}{Nature}

\newcommand{\solphys}{SoPh}

\newcommand{\Alf}{Alfv\'{e}n}

\newcommand{\beq}{\begin{equation}}
\newcommand{\eeq}{\end{equation}}

\DeclareMathAlphabet{\mathsfsl}{OT1}{cmss}{bx}{sl}
\SetMathAlphabet{\mathsfsl}{bold}{OT1}{cmss}{bx}{sl}

\usepackage{amssymb}
\usepackage{pifont}
\newcommand{\xmark}{\ding{55}}%

\usepackage{epstopdf}
\usepackage{color}        

\begin{document}

\title[striations: streamers or MHD waves?]
{Striations in molecular clouds: Streamers or MHD waves?}

\author[Tritsis and Tassis]
  {Aris~Tritsis$^{1}$ and Konstantinos~Tassis$^{1,2}$ \\
    $^1$Department of Physics and ITCP \thanks{Institute for Theoretical and Computational Physics, formerly Institute for Plasma Physics}, University of Crete, PO Box 2208, 71003 Heraklion, Greece\\
    $^2$IESL, Foundation for Research and Technology-Hellas, PO Box 1527, 71110 Heraklion, Crete, Greece}
\maketitle 

\begin{abstract}

Dust continuum and molecular observations of the low column density parts of molecular clouds have revealed the presence of elongated structures which appear to be well aligned with the magnetic field. These so-called striations are usually assumed to be streams that flow towards or away from denser regions. We perform ideal magnetohydrodynamic (MHD) simulations adopting four models that could account for the formation of such structures. In the first two models striations are created by velocity gradients between ambient, parallel streamlines along magnetic field lines. In the third model striations are formed as a result of a Kelvin-Helmholtz instability perpendicular to field lines. Finally, in the fourth model striations are formed from the nonlinear coupling of MHD waves due to density inhomogeneities. We assess the validity of each scenario by comparing the results from our simulations with previous observational studies and results obtained from the analysis of CO (J = 1 - 0) observations from the Taurus molecular cloud. We find that the first three models cannot reproduce the density contrast and the properties of the spatial power spectrum of a perpendicular cut to the long axes of striations. We conclude that the nonlinear coupling of MHD waves is the most probable formation mechanism of striations.

\end{abstract}


\section{Introduction}\label{intro}

Star formation occurs in condensations located within the dense elongated structures of molecular clouds. These structures, referred to as filaments, have been extensively studied both observationally and theoretically (see review of Andr{\'e} et al. 2014). Although the role of the magnetic field in the evolution of filaments is still a topic of debate, its topology with respect to these filaments is well established. Polarimetric studies (Moneti et al. 1984; Pereyra \& Magalh{\~a}es 2004; Alves et al. 2008; Chapman et al. 2011; Sugitani et al. 2011; Palmeirim et al. 2013; Planck Collaboration et al. 2014) have revealed that the magnetic field is well ordered near dense filaments and perpendicular to their long axis. 

Elongated structures, called striations, are also seen in the low column density parts of molecular clouds. Despite the fact that striations are not sites of star formation they are of high importance for interstellar medium (ISM) studies since they can reveal the dynamics of molecular clouds, and the early stages of star formation. However, there is yet no theoretically established physical mechanism explaining their formation. Understanding how striations form, whether they are long-lived or transient features and their role in star formation are important open questions.

Striations were first observed in $^{12}\rm{CO}$ and $^{13}\rm{CO}$ by Goldsmith et al. (2008) at the northwest part of the Taurus molecular cloud where they appear as autonomous structures. Striations were also observed by $\textit{Herschel}$ in dust emission. One of the most representative examples is the Polaris flare where well ordered, low density elongations are seen throughout the cloud (Miville-Desch{\^e}nes et al. 2010). Like in Taurus, striations in the Polaris flare do not appear to be associated with the denser parts of the cloud. However, in certain clouds, striations are connected to denser filaments. Hennemann et al. (2012), Palmeirim et al. (2013) and Alves de Oliveira et al. (2014) analysed $Herschel$ dust emission maps from DR21, Taurus and Chamaeleon molecular clouds respectively. In all of these studies, striations were interpreted as streamlines in which material flows into or out from more dense filaments and/or clumps. 

Malinen et al. (2015) compared $Herschel$ dust emission maps and Plank polarization data from the cloud L1642 in order to quantify the relative angle between the plane-of-the-sky component of the magnetic field and the long axes of striations. Using the Rolling Hough Transform (RHT) algorithm (Clark et al. 2014) they concluded that striations were in excellent alignment with the magnetic field. Panopoulou et al. (2016b) also used RHT to compare the orientation of the plane-of-the-sky (POS) magnetic field with linear structures in the Polaris flare and reported that the majority of striations were aligned with the projected magnetic field. The alignment between these structures and the magnetic field has also been pointed out in all clouds by all relevant studies in the literature (Goldsmith et al. 2008;  Chapman et al. 2011; Hennemann et al. 2012; Palmeirim et al. 2013; Alves de Oliveira et al. 2014).

Li et al. (2013) considered the overall morphology of the magnetic field with respect to both filaments and striations. They concluded that besides the formation of dense filaments from the gravitational contraction along field lines, strong magnetic fields could also act as the guiding channels of sub-\Alf ic flows, thus forming striations. In this mass-accretion/flows-along-field-lines paradigm, which is currently the most common interpretation of such structures, density fluctuations are presumably caused by pressure differences which are in turn caused by fluctuations of the streaming speed as expected by Bernoulli's principle.

A shear velocity between ambient streamlines would normally lead to a Kelvin-Helmholtz instability. However, the presence of the magnetic field can stabilize the flow as long as the velocity difference between ambient streamlines is less than two times the \Alf~speed (Frank et al. 1996). In an early theoretical work, Frank et al. (1996) performed 2D simulations in ideal magnetohydrodynamics (MHD) assuming super-\Alf ic velocities with opposite signs on either side of a shear layer and an initially ordered magnetic field. They showed that although a Kelvin-Helmholtz instability occurred early on, a stable, laminar flow was quickly developed due to the presence of the magnetic field. The final density configuration in their simulations was parallel elongated structures aligned with the magnetic field. 

Supersonic motions and other kinematic properties of molecular clouds have often been interpreted in terms of the presence of hydromagnetic waves (Arons \& Max 1975; Zweibel \& Josafatsson 1983). Specifically, the linewidth-size relation is attributed to \Alf~waves with long wavelengths and large amplitude (Mouschovias \& Psaltis 1995). These findings suggest that striations may also be connected to MHD waves.

In the present paper we explore four possible physical mechanisms that could create striations. Since flows along magnetic field lines have been proposed by previous observational studies and are currently considered to be the most plausible mechanism for the formation of striations, we explore two models involving such flows. In the first model, we assume a sub-\Alf ic bulk flow and sub-\Alf ic velocity gradients between ambient streamlines. In the second scenario, we repeat the super-\Alf ic simulations performed by Frank et al. (1996) by adopting values for the parameters involved appropriate for molecular cloud conditions. In the third model, sub-\Alf ic flows perpendicular to the magnetic field cause a Kelvin-Helmholtz instability which in turn produces striations. Finally, we consider an entirely different scenario in which striations are formed from the excitation of compressional magnetosonic waves. In this model, fluctuations of magnetic pressure create striations. Magnetosonic waves are naturally excited from \Alf~waves due to density inhomogeneities. The values adopted in our models are driven from observational results from the literature and analysis of observational data presented here. We find that in the first three models the density contrast between the linear structures that are formed is so low that essentially flows along or perpendicular to magnetic field lines fail to create striations. In contrast, the model including coupling of MHD waves successfully reproduces most of the observational properties of striations.

In section \S~\ref{observations} we quantify the observed properties of striations to facilitate a quantitative comparison to simulations. Numerical simulations of models involving streamers and corresponding results are described in \S~\ref{sub1}, \S~\ref{super} and \S~\ref{sub2}. In \S~\ref{ms_from_alf} we provide some theoretical background for our fourth physical model (MHD waves) and describe our results. We summarize and discuss our conclusions in \S~\ref{discuss}.

\section{Observations}\label{observations}

\begin{figure}
\includegraphics[width=1.0\columnwidth, clip]{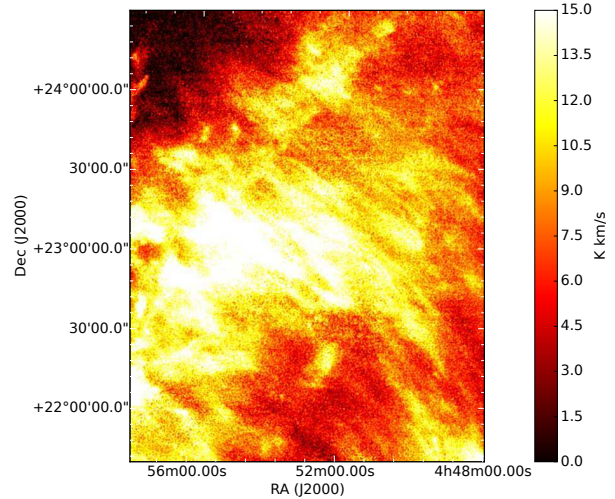}
\caption{CO integrated intensity map of the striations observed in Taurus molecular cloud. The integration was performed only in the velocity range where the striations appear (4.76 $\sim$ 7.55 km/s).
\label{str_zoomedin_map}}
\end{figure}

In order to observationally quantify the properties of the striations we use $^{12}$CO (J = 1 - 0) line emission data of the Taurus molecular cloud from the FCRAO survey (Goldsmith et al. 2008). The velocity resolution in the CO data cube is $\delta v_{ch}$= 0.266 km/s. FCRAO's telescope beam size at $^{12}$CO (J = 1 - 0) emission frequency (115 GHz) is 45" which at the distance of the Taurus cloud (140 pc) yields a spatial resolution of 0.013 pc. An integrated intensity map of the region of interest is shown in Figure~\ref{str_zoomedin_map}. 

\begin{figure}
\includegraphics[width=1.0\columnwidth, clip]{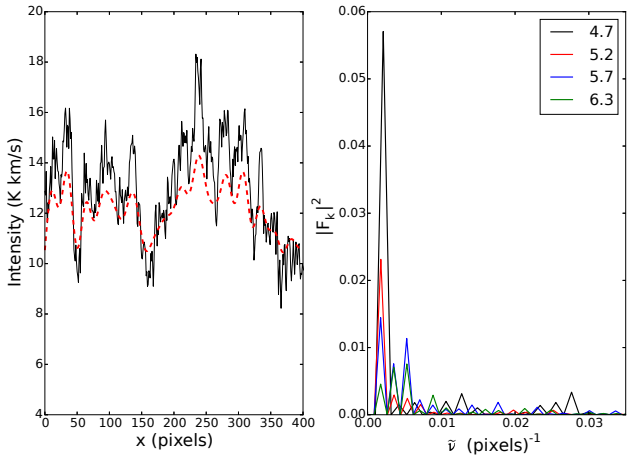}
\caption{Left panel: A cut perpendicular to the long axis of the striations (black line) and a low pass noise filter (smooth dashed red line). Right panel: The power spectrum in 4 velocity channels (given in km/s in the legend). All velocity channels exhibit the same dominant frequency with other wavelengths also present, thus suggesting a creation mechanism involving superposition of waves.
\label{vertcut_wnf_Fourierslices}}
\end{figure}

\begin{table*}
\centering
\caption{Parameters used in each run.}
\label{Models} 
\begin{tabular} {@{}c c c c c c c}
\hline\hline
 Model & density ($\rm{cm^{-3}}$) & $\rm{B_{y0}}$ ($\rm{\mu G}$) & $\delta$ (\%) & Chemistry & z dimension (pc) & resolution\\ 
\hline
\parbox[t]{0.8in}{sub-\Alf ic \\ streamers} & 200 & 15 & 10 & \xmark & \xmark & 128$\times$128\\
 &200 & 15 & 100 & \xmark & \xmark & 128$\times$128\\
 &200 & 15 & 100 & \xmark & \xmark & 256$\times$256\\
\hline
\parbox[t]{0.8in}{super-\Alf ic \\ streamers} & 200 & 15 & \xmark &\xmark & \xmark & 128$\times$128\\
 & 200 & 15 & \xmark & \xmark & 0.5 & 128$\times$128$\times$64\\
 & 200 & 15 & \xmark & \xmark & \xmark & 256$\times$256\\
\hline
\parbox[t]{0.8in}{sub-\Alf ic \\ flows $\perp$ to $\vec{B}$} & 200 & 15 & \xmark &\xmark & 0.5 & 128$\times$128$\times$64$^{\dagger\dagger}$ \\
\hline
\parbox[t]{0.8in}{MHD waves \\ coupling} & 200 & 15 & 15 & \checkmark & \xmark & 256$\times$256\\
 & 100 & 15 & 15 & \xmark & \xmark & 256$\times$256\\
 & 400 & 15 & 15 & \xmark & \xmark & 256$\times$256\\
 & 200 & 30 & 15 & \xmark & \xmark & 256$\times$256\\
 & 200 & 7.5& 15 & \xmark & \xmark & 256$\times$256\\
 & 200 & 15 & 30/15 & \xmark & \xmark & 256$\times$256$^{\dagger}$ \\
 & 200 & 15 & 30/15 & \xmark & \xmark & 256$\times$256$^{\ddagger}$ \\
 & 100 & 30 & 15 & \xmark & 0.125  & 256$\times$256$\times$64\\
 & 100 & 30 & 15 & \xmark & 0.25   & 256$\times$256$\times$64\\
 & 200 & 15 & 15 & \xmark & \xmark & 512$\times$512\\
\hline\hline
\multicolumn{7}{l}{$^{\dagger}$ The perturbation amplitude ($\delta$) is 30\% for the magnetic field and 15\% for the density and thermal pressure.}\\
\multicolumn{7}{l}{$^{\ddagger}$ Same as in $^{\dagger}$ but with a spectrum of \Alf~waves initially present in the system. }\\
\multicolumn{7}{l}{$^{\dagger\dagger}$ The value for the magnetic field strength refers to its $\textit{z}$ component. In this model, $\rm{B_{y0}}=\rm{B_{x0}}=0$.}\\
\end{tabular}
\end{table*}

In the observational data we define a cartesian coordinate system where the $\textit{x}$ and $\textit{y}$ axes are respectively perpendicular and parallel to the largest dimension of the striations as projected on the plane of the sky. Thus, the $\textit{z}$ axis is parallel to the line-of-sight (LOS). In the left panel of Figure~\ref{vertcut_wnf_Fourierslices} we show an averaged perpendicular cut of the integrated intensity of the striations (solid black line) to which we have applied a low pass filter (smooth dashed red line). In order to increase the signal to noise ratio in our analysis we first consider three adjacent cuts along the $\textit{x}$ direction and then average the intensity of their corresponding $\textit{y}$ pixel values. In the right panel of Figure~\ref{vertcut_wnf_Fourierslices} we show the spatial power spectra in 4 velocity channels. In the power spectrum there is unambiguously a dominant frequency with other wavelengths also present. To compute the power spectrum in each velocity slice we have again averaged pixel values from three adjacent cuts. The full velocity range where striations are visible is $\sim$ 2.5-3.0 km/s. Observations also suggest multiple velocity components along the LOS (Heyer \& Brunt 2012). From Figure~\ref{vertcut_wnf_Fourierslices} it is clear that both the integrated intensity cut and intensity cuts in velocity slices are quasi-periodic.

Due to the quasi-periodicity of the integrated intensity cut there is no unique contrast between maxima and minima. We thus need a method to robustly and systematically compute the contrast. To do so, we first consider all perpendicular cuts to the long axis of the striations and identify where each intensity cut has extrema. We compute the contrast between successive extrema and take the mean of all contrast values. To avoid confusion caused by point to point variations, extrema were identified from the low pass filter rather than the actual cut. However, the contrast was properly computed from the values of the actual intensity cut. The mean contrast, adopting a low pass filter such as the one shown in the left panel of Figure~\ref{vertcut_wnf_Fourierslices} (red dashed line) which reasonably follows the intensity profile, is $\sim$ 25\%. For direct comparison with observations we will use the same method of computing the contrast throughout this paper.

\section{Streamers}

We performed 2D and 3D numerical simulations in cartesian coordinates using the astrophysical code FLASH 4.0.1 (Fryxell et al. 2000; Dubey et al. 2008). We used the unsplit staggered mesh algorithm (Lee 2013) to solve the equations of ideal MHD without gravity. For the Riemann problem we used Roe's solver which accounts for all waves that can arise in the MHD equations. We used van Leer's flux limiter and third order interpolation to reduce numerical diffusion as much as possible.

In the simulations we adopted magnetic field and density values driven from observational estimates from the same region in Taurus. Chapman et al. (2011) used polarization data to map the POS component of the magnetic field at the northwest part of Taurus molecular cloud where striations were first observed. Using the Chandrasekhar \& Fermi (1953) method they found a value of $\rm{B_{pos}=17 \pm 1 ~\mu G}$ whereas using Hildebrand's et al. (2009) method they found $\rm{B_{pos}=31 \pm 4 ~\mu G}$. Despite the fact that the intrinsic magnetic field value would be even higher than these observational limits, we adopted a conservative reference value of $15 ~\rm{\mu G}$. Chapman et al. (2011) also used CO data from the FCRAO survey (Goldsmith et al. 2008) to constrain the number density. They reported a value of $\rm{\rho=200 \pm 10 ~cm^{-3}}$. This was the value used for the background number density in our reference runs. A constant temperature of 15 K was adopted for all of our models. Thus, the sound speed is $\sim$ 0.35 km/s, the \Alf ~speed ($v_a=B/\sqrt{4\pi \rho}$) is $\sim$ 1.58 km/s and the plasma $\beta$ parameter ($\beta=8\pi P_{th}/B^2$ where $P_{th}$ is the thermal pressure) is $\sim$ 0.1.

In our simulations, we define a cartesian coordinate system such that the direction of the magnetic field is along the $\textit{y}$ axis and the $\textit{z}$ axis represents the LOS dimension. The physical dimensions of the computational area in our 2D simulations are 1 pc in each direction. Driven from recent observational results (Qian et al. 2015), in 3D simulations, the LOS dimension is taken to be smaller than the other two. We terminate each simulation after 5 Myrs. A list of all runs is given in Table~\ref{Models}.
\begin{figure*}
\includegraphics[width=2.1\columnwidth, clip]{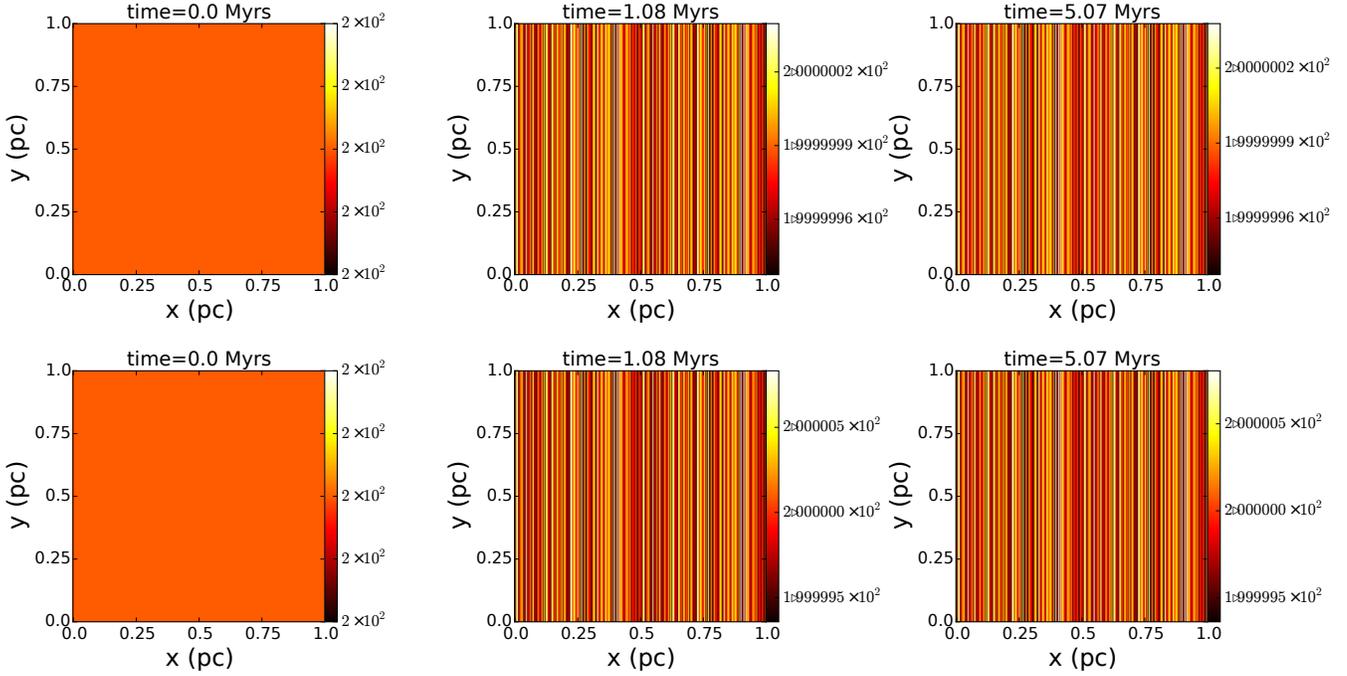}
\caption{Slice density maps from our 2D simulations for the sub-\Alf ic flow along field lines model. In the upper row we plot our results from the simulation where the amplitude of the perturbation is 10\% and in the lower row results where the amplitude of the perturbation is 100\%. From left to right column the time is 0 Myrs (i.e. the initial condition), $\sim$ 1 Myr and $\sim$ 5 Myrs. The contrast between adjacent striations is extremely low for both perturbation amplitudes.
\label{kh_superalfv_slices}}
\end{figure*}

\begin{figure*}
\includegraphics[width=2.0\columnwidth, clip]{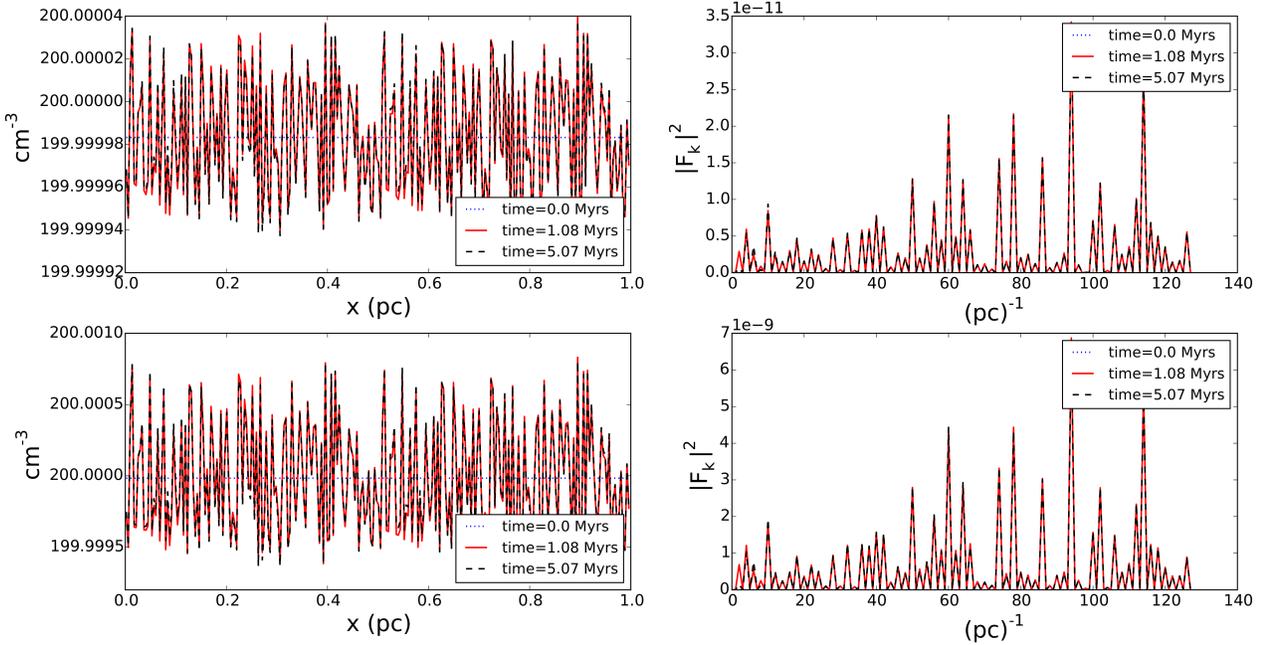}
\caption{Left column: perpendicular density cuts to the striations for 0 Myrs (blue dotted line), $\sim$ 1 Myr (solid red line) and $\sim$ 5 Myr (dashed black line) for the sub-\Alf ic flow along setup and for the two perturbation amplitudes (upper panel: 10\%, lower panel: 100\%). Right column: The corresponding spatial power spectra. 
\label{kh_subalfv_cuts_n_Fourier}}
\end{figure*}

\begin{figure*}
\includegraphics[width=2.0\columnwidth, clip]{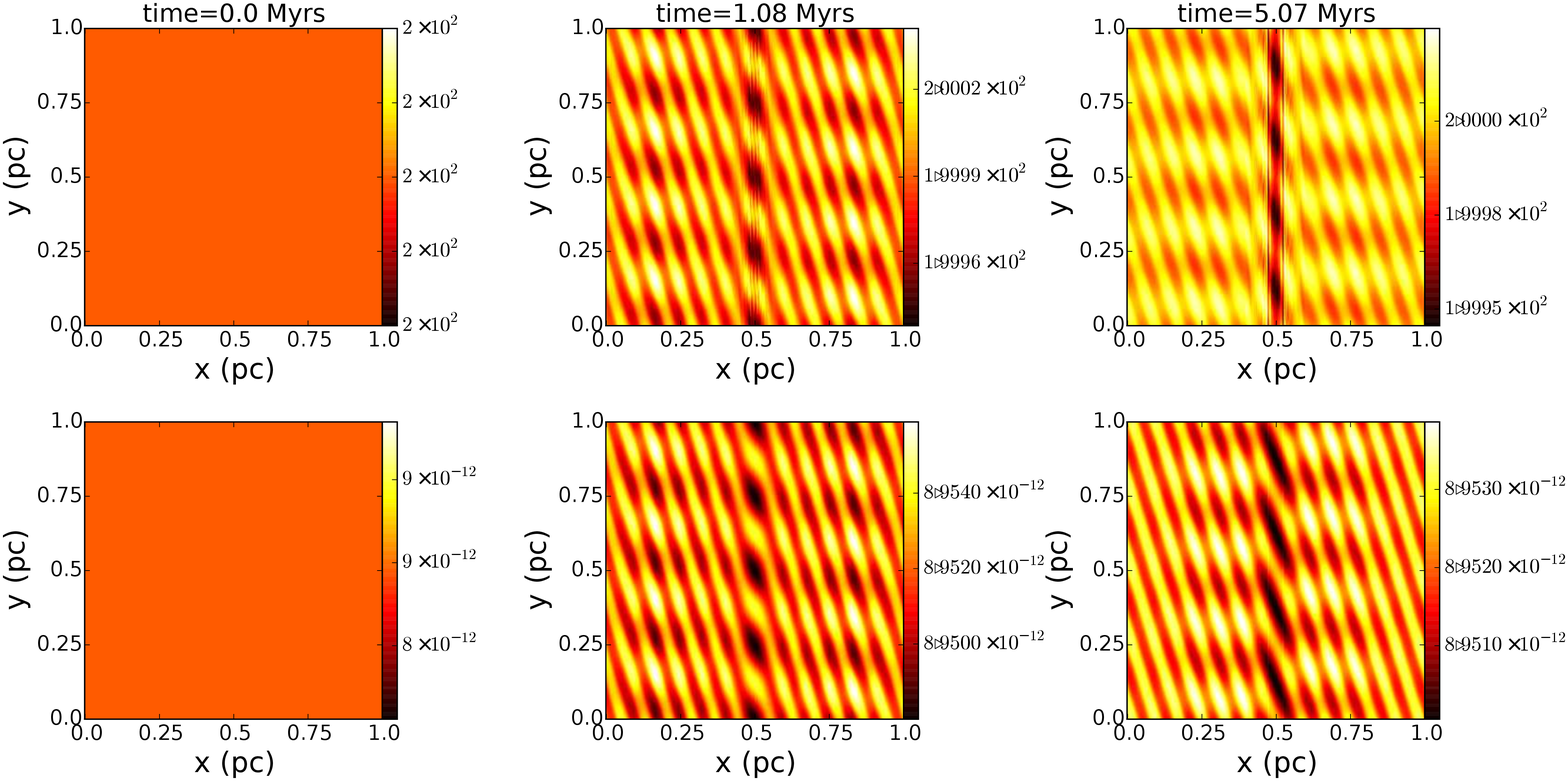}
\caption{Upper row: density maps for 0 Myrs (left), $\sim$ 1 Myr (middle) and $\sim$ 5 Myr (right) from the simulations with the super-\Alf ic flow along setup. Lower row: magnetic pressure maps for the same time steps. 
\label{superalfv_slices}}
\end{figure*} 
\subsection{sub-\Alf ic flow along field lines}\label{sub1}

In this model we test the premise that sub-\Alf ic velocity differences between ambient streamlines can form striations. In principle, velocity differences cause a pressure gradient in the perpendicular to the magnetic field direction which is in turn expected to create compressions and rarefactions in density. These shear flows are stabilized against Kelvin-Helmholtz instabilities due to the presence of the magnetic field. The stability condition for the Kelvin-Helmholtz instability assuming an inviscid, incompressible magnetized plasma is:
\begin{equation}\label{stability}
\frac{1}{4\pi}[(\vec{k}\cdot\vec{B_1})^2+(\vec{k}\cdot\vec{B_2})^2]>\frac{\rho_1\rho_2}{\rho_1+\rho_2}[\vec{k}\cdot(\vec{v_1}-\vec{v_2})]^2
\end{equation}
(Chandrasekhar 1961) where $\vec{\rm{k}}$ is the wavenumber, $\vec{\rm{B_i}}$ is the magnetic field value and $\vec{v_i}$ (i=1, 2) is the velocity on either side of the shear layer. In this sub-\Alf ic flow along field lines model, the stability condition of Equation~\ref{stability} is satisfied at all times.

Initially, we assume a sub-\Alf ic, constant flow along the direction of the magnetic field. We then introduce perturbations in the velocity field along the $\textit{x}$ direction. Thus, at the beginning of the simulation, the only non-zero component of the velocity is along the $\textit{y}$ axis and is given by:
\begin{equation}
v_y(x)=v_{y0}+\delta v_y(x)
\end{equation}
where $\delta v_y(x)$ is random and positive. We run simulations with two different amplitudes for the velocity perturbation, 10\% and 100\% of the unperturbed velocity. The unperturbed value of the $\textit{y}$ velocity component is 0.65 km/s. Thus the bulk velocity of the flow is sub-\Alf ic, yet supersonic. The magnetic field, density and thermal pressure are left unperturbed. 

The boundary conditions are periodic in the $\textit{y}$ axis (i.e. along field lines) and reflective in the $\textit{x}$ direction. Simulations for this model were performed on a uniform grid with 128$\times$128 grid points. Therefore, our spatial resolution is $\sim 7.8\times 10^{-3}$ pc. An additional simulation was performed on a 256 $\times$ 256 grid to ensure convergence (see Table~\ref{Models}). 

In Figure~\ref{kh_superalfv_slices} we show density maps for the simulations with 10\% (upper row) and 100\% (lower row) perturbation amplitudes in the y velocity component. Different columns represent different times. From left to right the time is 0 Myrs (i.e. the initial condition), $\sim$ 1Myr and $\sim$ 5 Myrs. The mean contrast in density between adjacent striations is extremely low. In fact the mean contrast in density for the simulation in which the velocity difference between ambient streamlines can be up to 100\% is just $\sim$ $\rm{4\times10^{-4}}$\%. The situation could not have been improved in a column density map if we had performed a 3D simulation. Projection effects from the 3D geometry would not increase the contrast since, due to the small size of the striation region in Taurus we can make the reasonable assumption that the LOS dimension has a constant thickness. As a result, the contrast in column density would roughly be the same as the contrast in density.

In Figure~\ref{kh_subalfv_cuts_n_Fourier} we show perpendicular cuts to the long axis of these streamlines (left column) and the spatial power spectrum (right column). We plot the results for the simulation with 10\% perturbation in the upper row and results for the simulation with 100\% perturbation in the lower row. The two power spectra have peaks in the same spatial frequencies. Neither the profiles nor the power spectra resemble observations, but instead they are consistent with white noise. The features seen in all panels of Figure~\ref{kh_subalfv_cuts_n_Fourier} originate from the random number generator with which we set up the velocity perturbations. 

Despite the fact that the $\textit{x}$ boundary conditions for this model are reflective no waves are excited. Instead, magnetic field lines are ``pushed" until the condition $\Pi_1=\Pi_2$ is satisfied in every interface between adjacent streamlines. Here, $\rm{\Pi=P+\frac{B^2}{8\pi}}$ is the total pressure. Furthermore, since the region is magnetically dominated, thermal pressure is not sufficient to cause large gradients in magnetic pressure in the $\textit{x}$ direction. As a result, $\rm{B_1\approx B_2}$, which for an isothermal gas leads to $\rm{\rho_1\approx \rho_2}$.


\subsection{super-\Alf ic flow along field lines}\label{super}

For the initial conditions in this model we partly follow Frank et al. (1996). The magnetic field is again directed along the $\textit{y}$ axis and the $\textit{y}$ component of the velocity is given by:
\begin{equation}\label{vel_field_super}
v_y=-v_0tanh(\frac{x-L_x/2}{a})
\end{equation}
where $v_0$ equals 1.2 times the \Alf~speed, $L_x$ is the size of the computational area in the $\textit{x}$ direction and $a$ is a parameter that quantifies the width of the shear layer. The value of $a$ is set at 4\% of the size of the $\textit{x}$ dimension. We further introduce a small velocity perturbation on the $\textit{x}$ component of the velocity with amplitude $10^{-3}$ the \Alf~speed. We do not perturb the magnetic field, thermal pressure or density. 

The boundary conditions are periodic in the $\textit{y}$ direction (i.e. along magnetic field lines) and outflow in the $\textit{x}$ direction. The 2D simulation for this model was performed on a 128$\times$128 uniform grid and an additional simulation with twice that resolution was performed in order to check for convergence. 

\begin{figure}
\includegraphics[width=1.\columnwidth, clip]{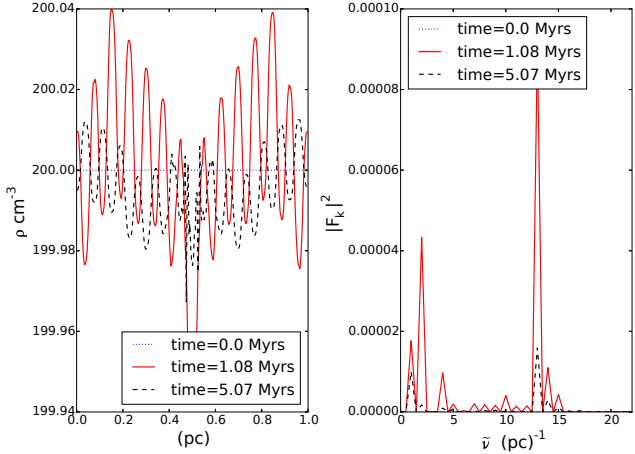}
\caption{Left panel: perpendicular density cuts to the striations for 0 Myrs (blue dotted line), $\sim$ 1 Myr (solid red line) and $\sim$ 5 Myr (dashed black line) for the super-\Alf ic streamers. Right panel: the power spectrum in the same three time steps. 
\label{kh_super_alfv_cut_n_Fourier}}
\end{figure}

In Figure~\ref{superalfv_slices} we show density maps (upper row) and magnetic pressure maps (lower row) for the same timesteps as in the previous model. As in the simulations by Frank et al.(1996) the flow is ordered due to the magnetic field despite the fact that the Kelvin-Helmholtz stability condition (Equation~\ref{stability}) is not satisfied. Intriguingly, the final configuration is parallel elongated structures which resemble observations. What is more, magnetic pressure fluctuations follow overdensities and rarefactions very well while magnetic field lines have been pivoted with respect to the initial configuration. Consequently, it seems that qualitatively this model may resemble observations. However, as in the case of the sub-\Alf ic streamers, the mean density contrast is very low, just $\sim$ $\rm{7.5\times10^{-3}}$\%.   

\begin{figure}
\includegraphics[width=1.0\columnwidth, clip]{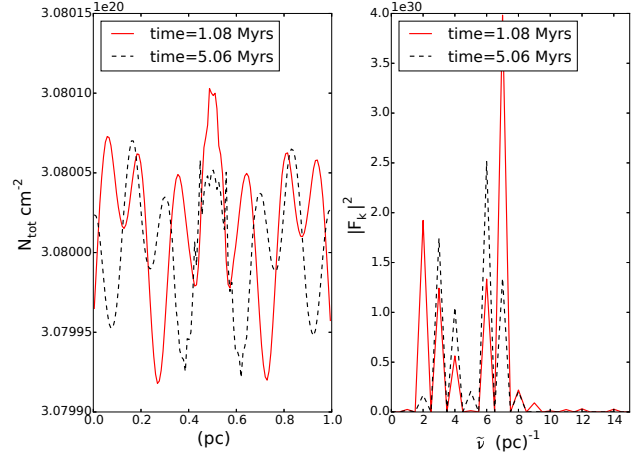}
\caption{Left panel: perpendicular column density cuts to the striations for $\sim$ 1 Myrs (solid red line) $\sim$ 5 Myrs (dashed black line) as projected along the $\textit{z}$ axis from our 3D simulation adopting the super-\Alf ic flow along field lines model. Right panel: The corresponding spatial power spectrum for the same two timesteps.
\label{kh_super3D_cuts_n_Fourier_on_axis_projections}}
\end{figure}

\begin{figure*}
\includegraphics[width=2.1\columnwidth, clip]{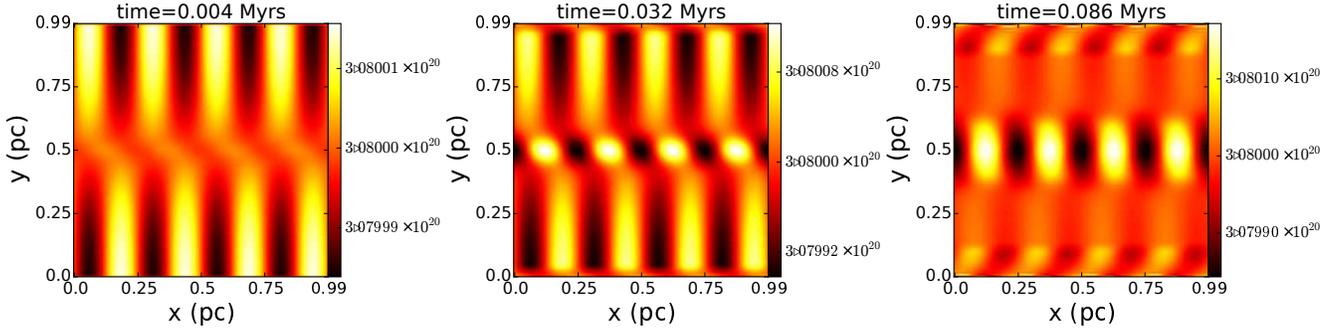}
\caption{Column density maps from our 3D simulations for the sub-\Alf ic flows perpendicular to the field lines model. From left to right the time is 0.004 Myrs (i.e. the initial condition), $\sim$ 0.03 Myr and $\sim$ 0.09 Myrs. As in the two previous models the contrast between adjacent striations is extremely low.
\label{kh_perp3D}}
\end{figure*}

In Figure~\ref{kh_super_alfv_cut_n_Fourier} we show perpendicular cuts for each of the timesteps of Figure~\ref{superalfv_slices} and the corresponding spatial power spectra. In contrast to sub-\Alf ic streamers, clear structures are created when the flow is super-\Alf ic. However, neither the perpndicular cut nor the power spectrum can reproduce observations. Here, power is distributed differently than observations in the sense that larger spatial frequencies have more power than the small ones. This is because the power is concentrated in the smallest ellipsoidal features seen in the density maps.

Even though the density contrast and power spectrum cannot reproduce observations, no reliable conclusions can yet be drawn from the 2D simulations alone since in three dimensions the magnetic field could affect the flow in a different manner. The possibility also exists that in 3D the magnetic field can no longer stabilize the flow, which would lead to turbulence. We thus run an additional 3D simulation where the shear layer is a sheet-like structure that extends along the LOS. The $\textit{y}$ component of the velocity is still given by Equation~\ref{vel_field_super}, but now, we also perturb $v_0$ along the $\textit{z}$ direction, i.e. $v_0(z)=v_0+\delta v(z)$. In 3D, in the $\textit{z}$ direction, we choose outflow boundary conditions, the physical length is 1/2 shorter than that of $\textit{x}$ and $\textit{y}$ and the resolution of the grid is 128$\times$128$\times$64.

In Figure~\ref{kh_super3D_cuts_n_Fourier_on_axis_projections} we show the results from our 3D simulation for $\sim$ 1 Myr and $\sim$ 5 Myrs. In the left panel we plot the profiles across projections along the $\textit{z}$ axis, and in the right panel the corresponding power spectra. Introducing a velocity profile along the $\textit{z}$-axis does not lead to turbulence but instead, the magnetic field can still stabilize the flow. However, as can be seen from the left panel of Figure~\ref{kh_super3D_cuts_n_Fourier_on_axis_projections} the contrast is still very low, even in a column density map. The low contrast remains regardless of the projection angle. In fact, when we consider a projection along the $\textit{z}$-axis which for the intended purposes of these simulations represents the LOS, the mean contrast is $\sim$ $\rm{7.8\times10^{-3}}$\%, just barely larger than the mean contrast in density maps.

Column density maps from our 3D simulations resemble the density maps from our 2D simulations shown in Figure~\ref{superalfv_slices} quite well. However, there are differences amongst the power spectra in 2D and 3D. These differences originate from two distinct effects. First, since in our 3D simulations the term $v_0$ in Equation~\ref{vel_field_super} is a function of the LOS, different structures are formed in different density slices along the $\textit{z}$ direction. Although, in average, these density slices are similar to the ones shown in Figure~\ref{superalfv_slices} there are deviations which will appear in the power spectrum. The second reason is due to the integration along the LOS. In a column density map the bulges seen in the upper panel of Figure~\ref{superalfv_slices} are enhanced whereas other features are even fainter. Finally, in 3D, magnetic field lines are pivoted to a larger angle with respect to the initial configuration. In our 2D simulations the angle between the magnetic field at later times and the initial magnetic field is $\sim$ 18$^\circ$ whereas the same angle in our 3D simulations is $\sim$ 40$^\circ$.

From the stability condition of Equation~\ref{stability} when the density and magnetic field values are equal on either side of an interface, a Kelvin-Helmholtz instability occurs only if the velocity difference between the two layers is twice the \Alf~ speed. However, no such extreme velocity gradients have ever been observed in molecular clouds. Heyer \& Brunt (2012) did a careful analysis of the velocity field in Taurus using the same observational data used here and concluded that in the striations region the motions of the flow were trans-\Alf ic. As a result, such a mechanism of producing striations in the diffuse ISM, requiring so large velocity gradients, would be unphysical even if it could reproduce observations.

\subsection{sub-\Alf ic flow perpendicular to field lines}\label{sub2}

Periodically spaced, elongated structures, referred to as Ripples, have also been observed in the south-west part of Orion molecular cloud (Bern{\'e} et al. 2010). Bern{\'e} et al. 2010 attributed the formation of Ripples to a Kelvin-Helmholtz instability. Bern{\'e} \& Matsumoto (2012) and Hendrix et al. (2015) performed 2D and 3D MHD simulations respectively with flows perpendicular to the magnetic field and with the shear layer parallel to the field lines. They found that with this configuration periodically spaced, elongated structures were created. Although the physical conditions in these Ripples are very different from the physical conditions in regions where striations are observed, it is possible that the same formation mechanism is in operation. The possibility that Ripples and striations are created through the same physical process has been recently pointed out by Heyer et al. (2016).

In order to test for this scenario, we performed a 3D simulation adopting the same values for the density and magnetic field strength as in the previous two models. The direction of the magnetic field was taken to be along the LOS (i.e. along the $\textit{z}$ axis) and the $\textit{x}$ component of the velocity was given by: 
\begin{equation}\label{vel_field_subperp}
v_x=-v_0tanh(\frac{y-L_y/2}{a})
\end{equation}
where similarly to Equation ~\ref{vel_field_super} the width of the shear layer is taken to be 4\% the size of the $\textit{y}$ dimension. In equivalence with our 3D super-\Alf ic simulations $v_0$ was also perturbed along the $\textit{z}$ axis. With this configuration, the shear layer extends along the LOS and is thus parallel to the magnetic field.

Super-\Alf ic flows perpendicular to field lines would lead to distortions of the magnetic field. However, the magnetic field in the regions where striations appear is well ordered. Thus, such flows must be sub-\Alf ic and the value of $v_0$ in Equation~\ref{vel_field_subperp} was taken to equal 0.45 times the \Alf ~speed. Small amplitude perturbations were further introduced in the $\textit{y}$ velocity component. Apart from the velocity field, all other quantities were left unperturbed. Boundaries were periodic in all directions, the length of the LOS was half that of the other two directions and the resolution of the grid was 128$\times$128$\times$64.

In Figure~\ref{kh_perp3D} we plot column density maps for 3 timesteps, early in the simulation before turbulence is developed. Similarly to the two previous models involving flows, the contrast between adjacent striations is extremely small. Applying the same method as in observations in the column density map shown in the right panel of Figure~\ref{kh_perp3D} results in a mean contrast of just 0.004\%.

In this model, the direction of the magnetic field is perpendicular to the planes shown in Figure~\ref{kh_perp3D}. As a result, this viewing angle would not result in elongated structures parallel to the magnetic field since the observed polarization intensity would be zero. Moreover, if the magnetic field is at an angle with respect to the $\textit{z}$ axis the growth of the instability would decrease or even halted for large angles. Thus, there is a limited range of viewing angles for which the depolarization factor due to projection effects is small and these structures appear as elongations. Furthermore, elongations formed via this mechanism are only transient features lasting just a few $\rm{10^4}$ years. At later times, turbulence develops and these structures are no longer recognizable. Consequently there are at least two additional shortcomings in terms of matching the observations.

\section{MHD waves}\label{ms_from_alf}

\begin{figure}
\includegraphics[width=1.0\columnwidth, clip]{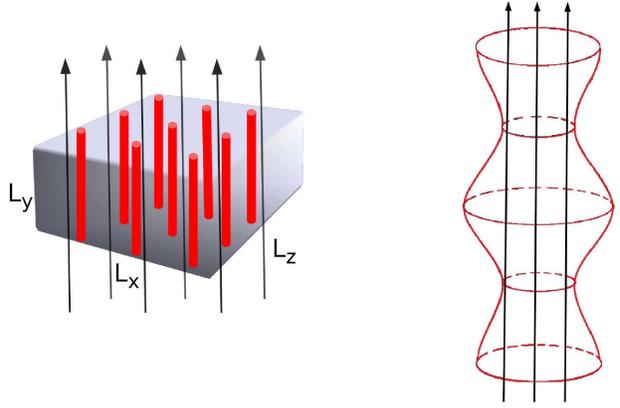}
\caption{Left panel: cartoon representation of the final configuration from the excitation of magnetosonic waves in both directions perpendicular to the magnetic field. The red cylinders represent density enhancements. Right panel: sausage waves in each cylindrical-like structure of the left panel. Black arrows denote the direction of the ordered magnetic field and the red line represents the morphology of the resulting magnetic flux tube due to perturbations.
\label{model}}
\end{figure} 

\begin{figure*}
\includegraphics[width=2.1\columnwidth, clip]{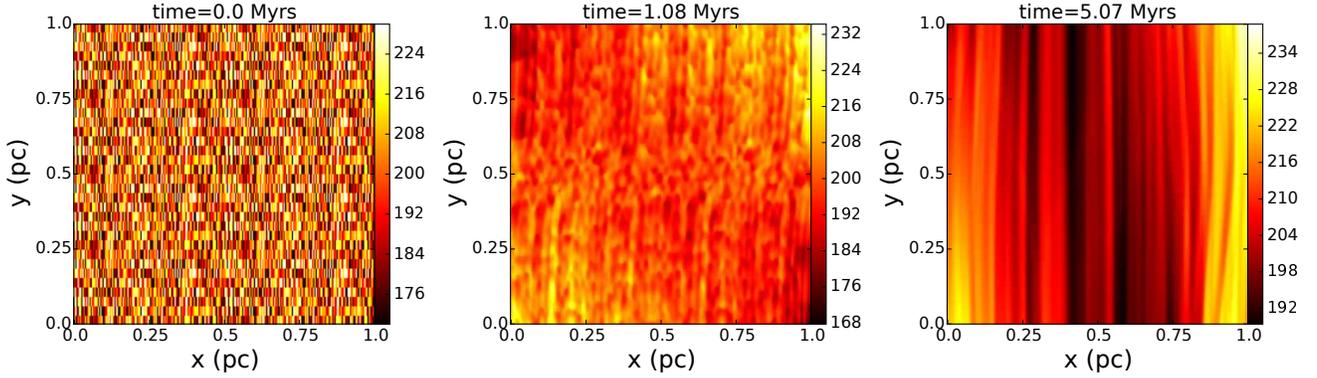}
\caption{Slice density maps from our 2D simulations for the fourth model. From left to right column the time is 0 Myrs (i.e. the initial condition), $\sim$ 1Myr and $\sim$ 5 Myrs. The contrast is drastically increased compared with the previous three models.
\label{ms_from_Alfven_default_2Dmaps}}
\end{figure*} 

The equations of ideal MHD can be linearised by considering small amplitude perturbations. If we then assume that the perturbed quantities vary as $\rm{e^{i(\vec{k}\cdot\vec{r}-\omega t)}}$ and substitute the expressions that arise for the perturbed quantities in the linearised equation of motion we get:
\begin{eqnarray}\label{lin_mom_eq}
&& [\omega^2-\frac{(\vec{k}\cdot\vec{B_0})^2}{4\pi\rho_0}]\vec{v}=\\ \nonumber
&& \{[c_s^2+\frac{B_0^2}{4\pi\rho_0}]\vec{k}-\frac{(\vec{k}\cdot\vec{B_0})}{4\pi\rho_0}\vec{B_0}\}(\vec{k}\cdot\vec{v})-\frac{(\vec{k}\cdot\vec{B_0})(\vec{v}\cdot\vec{B_0})}{4\pi\rho_0}\vec{k}
\end{eqnarray} 
where $\vec{k}$ is the wavenumber, $\vec{B_0}$ is the unperturbed magnetic field, $\rho_0$ is the unperturbed density, $\omega$ is the angular frequency and $\vec{v}$ is the perturbed velocity. When the direction of propagation of the waves is parallel to the magnetic field (i.e. $\vec{k}$ $\parallel$ $\vec{B}$), Equation~\ref{lin_mom_eq} leads to the dispersion relation of \Alf~waves. In the situation with $\vec{k}$ $\bot$ $\vec{B}$, Equation~\ref{lin_mom_eq} leads to the dispersion relation of compressive magnetosonic waves which however, in the general case, can propagate at other angles as well.

In the linear regime \Alf~and fast magnetosonic waves propagate independently. However, when nonlinear terms are non-negligible, plasma density inhomogeneities across the direction of the magnetic field lead to phase mixing of \Alf~waves (Heyvaerts \& Priest 1983). As a result, fast magnetosonic waves can be excited even if they are not originally present in the system. Due to density inhomogeneities, there are also regions of varying \Alf~speed where fast magnetosonic waves get refracted and thus, they naturally get trapped inside overdensities or, in other words, in regions of low \Alf~speed.

We now consider an \Alf~wave of the form $Acos\omega(t-y/v_a(x))$, where $v_a$ is the \Alf~speed and $A$ is the amplitude of the wave, travelling along the direction of the unperturbed magnetic field which, as in the previous models, we assume is directed along the $\textit{y}$ axis. Then, the wave equation for fast magnetosonic waves can be written as:
\begin{eqnarray}\label{wave_magnetosonic}
\frac{\partial^2 v_x}{\partial t^2}-v_a^2(x)\Big(\frac{\partial^2 v_x}{\partial x^2}+ \frac{\partial^2 v_x}{\partial y^2} \Big)= \frac{\omega A^2}{v_a^2}\frac{dv_a(x)}{dx} \times\nonumber \\ 
\times\Big[\omega ycos\big(2\omega(t-\frac{y}{v_a(x)})\big)-v_a(x)sin\big(2\omega(t-\frac{y}{v_a(x)})\big) \Big]
\end{eqnarray}
where we have ignored variations along the $\textit{z}$ direction. From Equation~\ref{wave_magnetosonic} it can be seen that magnetosonic waves produced due to phase mixing will have twice the wavelength of \Alf~waves. Finally, fast magnetosonic waves travelling across magnetic field lines can get further refracted at the edges of the cloud, at sharp density gradients. For an extensive analytical discussion of the coupling between \Alf~and fast magnetosonic waves we refer the reader to Nakariakov et al. (1997). 

When the direction of propagation of the waves is perpendicular to the magnetic field ($\vec{k}$ $\bot$ $\vec{B}$), Equation~\ref{lin_mom_eq} can be written in component form as:
\begin{equation}
\omega^2v_x=(c_s^2+v_a^2)(k_xv_x+k_zv_z)k_x
\end{equation}  
\begin{equation}
\omega^2v_y=0
\end{equation}  
\begin{equation}
\omega^2v_z=(c_s^2+v_a^2)(k_xv_x+k_zv_z)k_z
\end{equation}  
where $\rm{c_s}$ is the sound speed. Consequently, magnetosonic waves travelling in both directions perpendicular to the magnetic field are also coupled to each other. In 3D, the final configuration will be cylindrical-like structures parallel to the magnetic field (left panel of Figure~\ref{model}). 

From the linearized MHD equations in cylindrical coordinates it follows that:
\begin{equation}\label{cylindrical1}
B_{c}=\frac{B_{0c}}{\omega_c}\frac{1}{r_c}\frac{\partial (r_cA(r_c))}{\partial r_c}sin(k_cz-\omega_c t)
\end{equation}
\begin{equation}\label{cylindrical2}
P_c=-\frac{\omega_c\rho_{0c}c_s^2}{(c_s^2k_c^2-\omega_c)}\frac{\partial (r_cA(r_c))}{\partial r_c}sin(k_cz-\omega_c t)
\end{equation}
(Freij et al. 2016). Here, $r_c$ is the radius of each cylindrical-like structure, $A(r_c)$ is the area and $P_c$, and $B_c$ are the perturbed pressure and magnetic field respectively.  As a result, for a cylindrical-like structure, the magnetic pressure and thermal pressure will be out of phase. This excites sausage waves (right panel of Figure~\ref{model}) along with the other two MHD wave modes.

From the right panel of Figure~\ref{model} it can be seen that the cross-sectional area of each flux tube changes. Thus, from Bernoulli's principle, there will be a velocity gradient along the long axis of each streamline. Ignoring gravity and assuming isothermal processes arises:
\begin{equation}\label{bernoulli}
(v_{1}^2-v_{2}^2)=2c_s^2ln(\frac{\rho_2}{\rho_1})
\end{equation} 
where the subscripts denote the velocity and density in different positions along one streamline. Although this is a second order effect, we should expect density and velocity fluctuations along the long axis of the striations.

In order to test if the coupling of \Alf~and fast magnetosonic waves can lead to the formation of striations we perturb the $\textit{x}$ component of the magnetic field while the ordered component is again towards the $\textit{y}$ direction. Additionally, we introduce random perturbations in density and thermal pressure in all directions in a self-consistent manner such that isothermality is never violated. All velocity components are initially set to zero. The initial conditions for this model are:
\begin{equation}
v_x=v_y=v_z=0
\end{equation}
\begin{equation}
\rho(x, z)=\rho_0+\delta\rho(x, y, z)
\end{equation}
\begin{equation}
P(x, z)=P_0+\delta P(x, y, z)
\end{equation}
\begin{equation}
B_y(x, z)=B_{0} 
\end{equation}
\begin{equation}
B_x(y)=B_z(y)=\delta Bsin(k_yy)
\end{equation}
where $k_y=\pi/L_y$. This setup implies an \Alf~wave passing through the computational region with wavelength twice the length of the simulated region in the $\textit{y}$ direction. For these simulations we used a fixed resolution grid with 256$\times$256 cells. Since matter can flow easily along magnetic field lines, the boundary conditions along the direction of the ordered component of the magnetic field are outflow. On the other hand, since magnetosonic waves can get reflected at the edges of the cloud, we use reflective boundaries in the $\textit{x}$ direction. 
\begin{figure}
\includegraphics[width=1.0\columnwidth, clip]{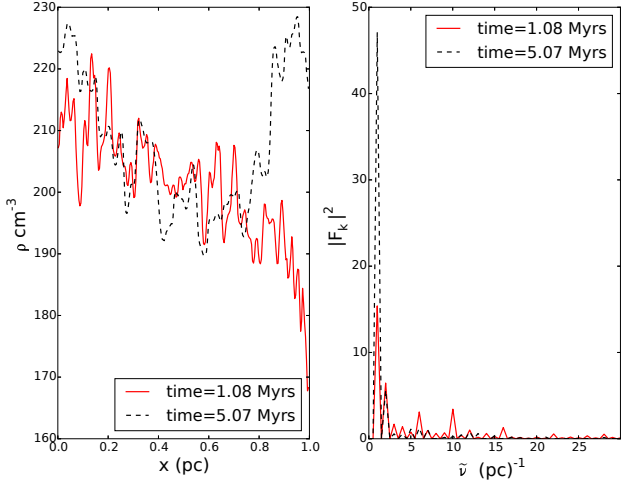}
\caption{Perpendicular density cuts (left panel) and the spatial power spectrum (right panel) for the timesteps shown in the middle and right panel of Figure~\ref{ms_from_Alfven_default_2Dmaps}. The power spectrum resembles observations extremely well.
\label{ms_from_Alfven_cuts_n_Fourier}}
\end{figure}

Using Hildebrand's et al. (2009) method, a number of authors (e.g. Eswaraiah et al. 2013; Franco \& Alves 2015) have found that the ratio of the random component of the magnetic field to the ordered component is of the order of 10\% and can be up to 17\%. Likewise, using the same method and the results presented in Chapman et al. (2011) it is found that in the northwest part of the Taurus molecular cloud the ratio of the turbulent to the ordered component of the magnetic field is $\sim$ 11\%. In general, the ratio of the ordered to turbulent component in molecular clouds can be up to $\sim$ 75\% (Houde et al. 2016) and, within uncertainties, it can also be up to 40\% in regions where striations appear (Panopoulou et al. 2016b). For the strength of the magnetic field in the striations region in the Polaris flare, Panopoulou et al. (2016b) reported values that ranged up to $\sim$ 80 $\mu$G. Since striations appear in regions of well ordered magnetic field our reference value for the amplitude of the perturbation is 15\% the background value of each perturbed quantity. However, we further explore how our results depend on this parameter by performing a run in which the ratio of the turbulent to ordered component of the magnetic field is 30\%. In this run, the amplitude of the perturbation for the density and thermal pressure is still 15\% and reference values were adopted for the density and the magnetic field.



\begin{figure}
\includegraphics[width=1.0\columnwidth, clip]{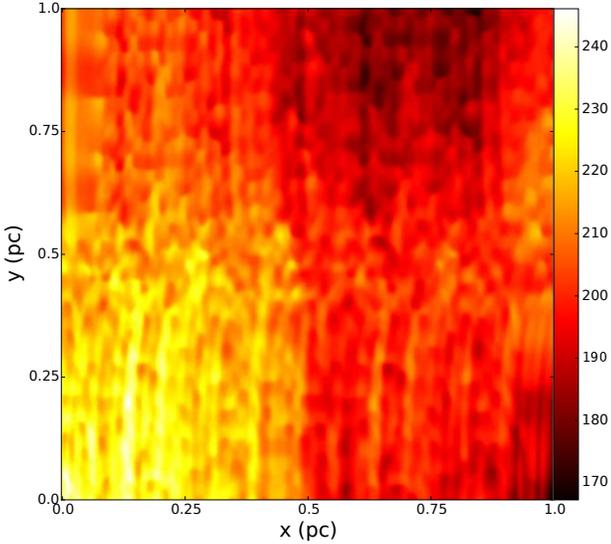}
\caption{Slice density map from our 2D simulation for the fourth model and for 1 Myr adopting a higher perturbation amplitude of 30\%. 
\label{ms_from_Alfven_default_2Dmaps_higher_d}}
\end{figure} 

\begin{figure}
\includegraphics[width=1.0\columnwidth, clip]{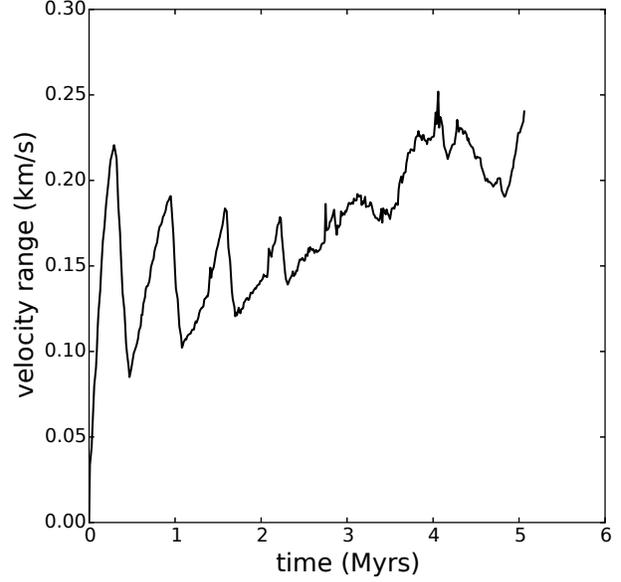}
\caption{Evolution of the velocity range in the direction perpendicular to the unperturbed magnetic field in the simulation shown in Figure~\ref{ms_from_Alfven_default_2Dmaps_higher_d}. No significant energy decay is seen in this simulation.}
\label{velocity_range_evolution}
\end{figure}

Density maps for the fourth model and for our reference run are shown in Figure~\ref{ms_from_Alfven_default_2Dmaps}. Not only can this model qualitatively reproduce observations but the contrast is drastically improved compared to the previous models. In fact, following the exact same procedure as in observations where we compute the contrast between adjacent maxima and minima for all perpendicular cuts, the mean contrast is $\sim$ 4.8\%. A noticeable feature in the middle panel of Figure~\ref{ms_from_Alfven_default_2Dmaps} is the mirror symmetry between the upper and lower half of the computational region. The reason behind this symmetry is that the derivatives in the nonlinear terms above and bellow the line $y=L_y/2$ appear in opposite signs. Furthermore, the region close to the line $y=L_y/2$ is where phase mixing mainly occurs. Then magnetosonic waves travel towards the lower and upper parts where the \Alf~speed is lower. These features are not so prominent in the right panel due to the outflow boundary conditions. After magnetosonic waves get excited they propagate in all directions, not just perpendicularly to magnetic field lines, and thus escape the computation region. This effect smoothens sharp density gradients and eliminates the mirror symmetry. We find that numerical diffusion does not significantly affect our results (see Apendix~\ref{convergence}).

\begin{figure}
\includegraphics[width=1.0\columnwidth, clip]{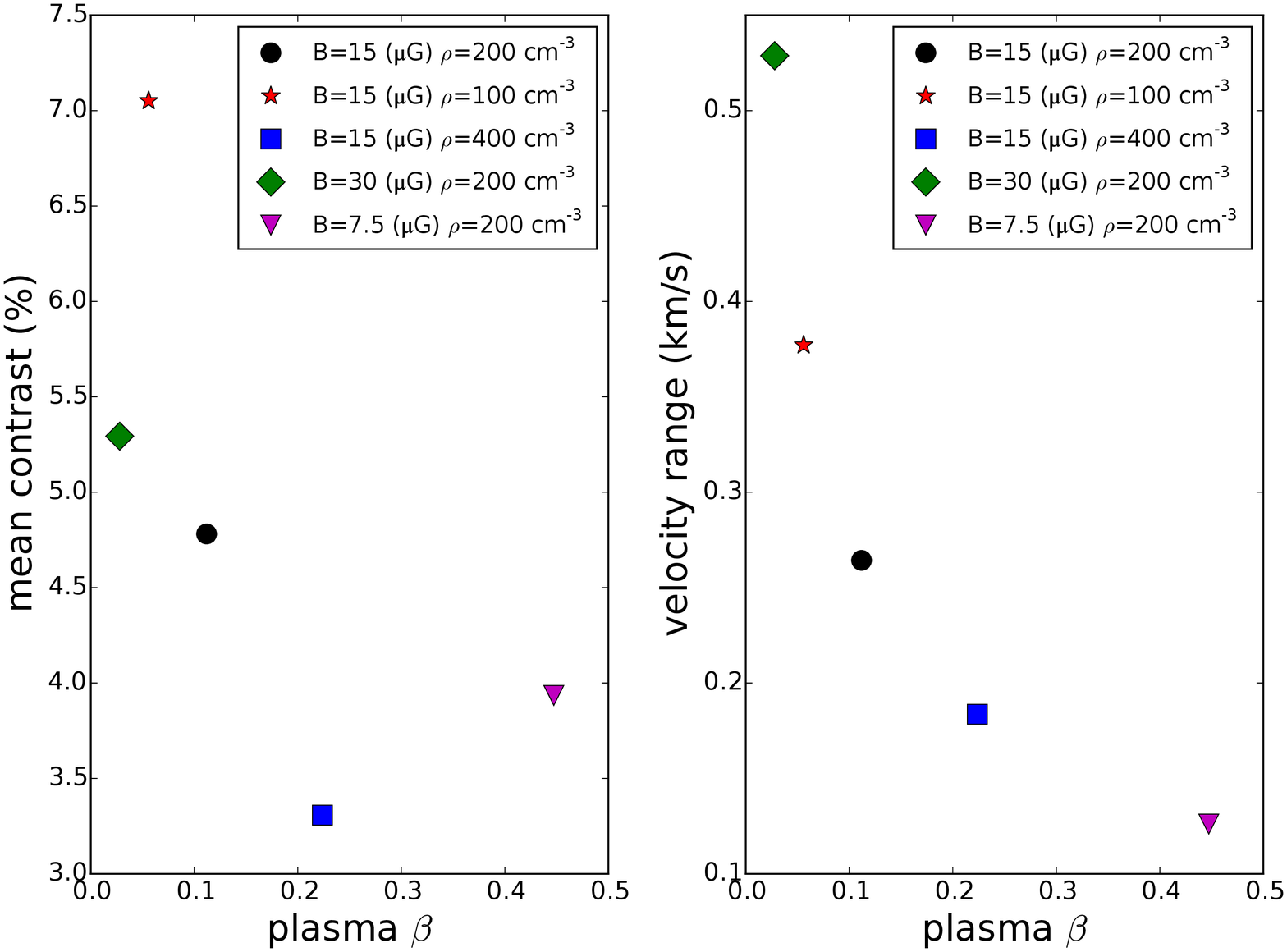}
\caption{Left panel: mean contrast as a function of plasma $\rm{\beta}$ for our fourth  model. The time in all parameter runs is $\sim$ 5 Myrs. In the simulations with lower plasma $\rm{\beta}$ the contrast is larger. Right panel: Velocity range as a function of plasma $\rm{\beta}$ for our fourth  model. The lower the plasma $\rm{\beta}$ the greater the velocity range reached throughout the simulation.}
\label{parameter_space}
\end{figure} 

In Figure~\ref{ms_from_Alfven_cuts_n_Fourier} we show perpendicular density cuts (left panel) and the spatial power spectrum (right panel). The solid red and dashed black lines correspond to the middle and right panel of Figure~\ref{ms_from_Alfven_default_2Dmaps} respectively. The power spectrum resembles observations remarkably well. Most of the power is distributed in larger wavelengths and, like observations, there are smaller peaks at larger spatial frequencies. These peaks correspond to the thinner structures seen in the middle and left panels of Figure~\ref{ms_from_Alfven_default_2Dmaps}. In agreement with the analytical result from Equation~\ref{wave_magnetosonic}, the dominant frequency in the right panel of Figure~\ref{ms_from_Alfven_cuts_n_Fourier} corresponds to a magnetosonic wave with wavelength two times that of the \Alf~wave initially present in the system. 

A density map from our simulation in which the amplitude of the perturbation is 30\% is shown in Figure~\ref{ms_from_Alfven_default_2Dmaps_higher_d}. The mean contrast in this simulation is 7.7\%. In Figure~\ref{velocity_range_evolution} we show the evolution of the maximum range of the $\textit{x}$ velocity component for 5 Myrs from the same run. When the amplitude of the \Alf~wave initially present in the system is large compared to the ordered component there is no significant energy decay. Basu \& Dapp (2010) were the first to report long-lived MHD modes without any dissipation in their simulations. The evolution of the maximum range of the $\textit{x}$ velocity component for the simulation with lower perturbation amplitude is given in Appendix~\ref{convergence}.

Since for this model results adopting reference parameters were in fairly good agreement with observations we additionally explored how the density contrast and velocity range changed by altering the initial density and the magnetic field strength a factor of two above and bellow the reference values. Furthermore, since striations are also observed in $\rm{CO}$ emission lines we couple one of our 2D runs with a non-equilibrium chemical model and investigated the correlation between $\rm{CO}$ abundance and the total density. Our chemical network consists of 13967 reactions that govern the evolution of 214 gas-phase species and 82 dust grain species. For the chemical modelling we assume a mean molecular weight of $\sim$ 2.4, a standard value of $\rm{\zeta=1.3 \times 10^{-17} ~s^{-1}}$ for the cosmic-ray ionization rate, the visual extinction is $\rm{A_v=1}$ mag and the temperature is constant and equal to 15 K. A list of the species included in the chemical network and values for the initial elemental abundances are given in Tritsis et al. (2016). 

\begin{figure}
\includegraphics[width=1.0\columnwidth, clip]{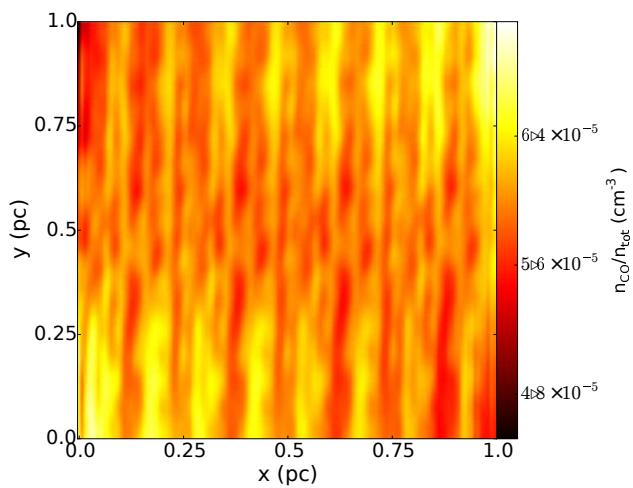}
\caption{CO abundance map for 1 Myr adopting our fourth model. In comparison to the middle panel of Figure~\ref{ms_from_Alfven_default_2Dmaps} striations are more prominent when seen through the chemical lens.}
\label{COmap}
\end{figure}

\begin{figure}
\includegraphics[width=1.0\columnwidth, clip]{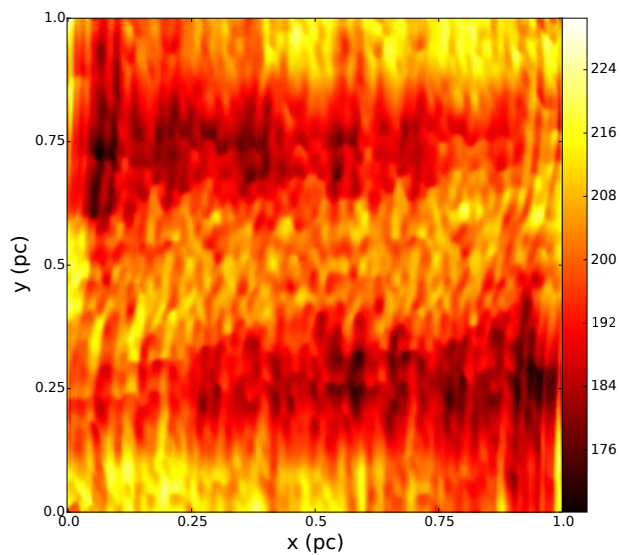}
\caption{Slice density map from our 2D simulation with 30\% perturbation amplitude and a spectrum of \Alf~waves initially present in the system. The time is 1 Myr. The final density configuration is eve more realistic.}
\label{spectrum_of_Alf}
\end{figure}

In the left panel of Figure~\ref{parameter_space} we show the mean contrast as a function of the mean plasma $\rm{\beta}$ for a time of 5 Myrs. The contrast has again been computed in the exact same manner as in observations. We find that the contrast is affected by both the value of the magnetic field and density independently, although the highest values are obtained for low plasma $\rm{\beta}$. In the right panel of Figure~\ref{parameter_space} we plot the maximum velocity range reached throughout each simulation in our parameter study as a function of the mean plasma $\rm{\beta}$. The lower the plasma $\rm{\beta}$, the greater the velocity range. The velocity range has a clear relation with plasma $\rm{\beta}$. In fact, it scales as $\sim$ $1/\sqrt{\rm{\beta}}$. 

In Figure~\ref{COmap} we show a $\rm{CO}$ abundance map from the 2D simulation in which we have included chemical modelling. The time is 1 Myr and thus it corresponds to the middle panel of Figure~\ref{ms_from_Alfven_default_2Dmaps}. Striations seen in the $\rm{CO}$ abundance map are more prominent than the ones seen in the total density map. The differences seen in the two maps are not just qualitative. The mean contrast that arises from the $\rm{CO}$ abundance map is 6.3 \% in comparison to 4.8 \% from the density map. The reason behind this discrepancy is the difference between dynamical and chemical timescales.

In the setup adopted so far for the 2D simulations we only considered one \Alf~wave passing through the computational area with wavelength two times the length of the $\textit{y}$ direction. In nature, we should expect a spectrum of \Alf~waves. Smaller and sharper distortions of the magnetic field would lead to larger gradients which would in turn make non-linear terms even more significant and lead to larger velocity ranges. We thus performed an additional 2D simulation where a superposition of three \Alf~waves with random phases was initially present in the system. The total amplitude of the perturbation in this simulation was 30\% with the amplitude of each \Alf~wave decreasing with wavelength. Specifically, the ratio of amplitudes of the longest and intermediate wavelengths with respect to the shortest wavelength was three and two respectively. A density map from this simulation is shown in Figure~\ref{spectrum_of_Alf}. The final configuration is even more realistic and the similarity with dust continuum observations of the striations region in the Polaris flare (Miville-Desch{\^e}nes et al. 2010) is remarkable. The maximum velocity range along the \textit{x} direction achieved throughout this simulation was $\sim$0.53 km/s, i.e more than a factor of two larger than in the simulation with the same total perturbation amplitude and just one \Alf~wave.

In order to examine the properties of striations in velocity slices we run additional 3D simulations of the same model considering two different lengths for the LOS dimension, 0.25 pc and 0.125 pc. In these simulations we adopt values for the unperturbed density and magnetic field that best agree with observations as these have arisen from the parameter study in our 2D simulations. The unperturbed magnetic field is $30 ~\rm{\mu G}$ and the number density is set at $\rm{100 ~cm^{-3}}$. These values are still well within observational limits. We also perturb the $\textit{z}$ component of the magnetic field along with the $\textit{x}$ component. In equivalence to our reference 2D simulation of the same model we only consider one \Alf~wave with wavelength twice the length of the $\textit{y}$ direction. The boundary conditions along the $\textit{z}$ axis are reflective and along the other two directions are kept as in our 2D simulations.

\begin{figure}
\includegraphics[width=1.0\columnwidth, clip]{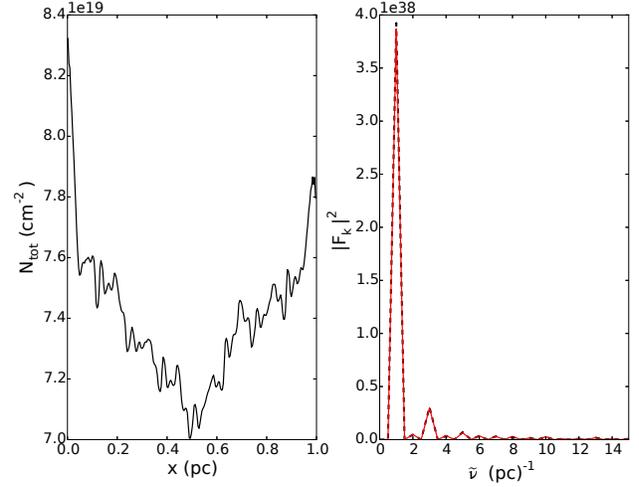}
\caption{Left panel: A perpendicular cut to the long axis of the striations. The cut has arose from a column density map from the simulation where $\textit{z}$=0.25 pc and when the time is 0.95 Myrs. Right panel: spatial power spectra of two velocity slices. The velocity slices are $\sim$ 0.2 km/s apart. In absolute agreement with observations the two velocity slices exhibit the same spatial frequencies.}
\label{column_denscut_vel_slice_fourier}
\end{figure}

The velocity range along the LOS in both our 3D simulations is $\sim$ 0.9 km/s. In the simulation where the LOS dimension is 1/4 that of the other two the maximum velocity range is obtained for $\sim$ 0.95 Myrs whereas when the LOS is even shorter the maximum range is achieved later on during the evolution. We find that the simulation where the length of the $\textit{z}$ dimension is 0.25 pc can better reproduce observations. In the left panel of Figure~\ref{column_denscut_vel_slice_fourier} we show a perpendicular cut from a column density map. In the right panel we plot the spatial power spectra of two velocity slices which are 0.2 km/s apart. In complete analogy with observations the two slices exhibit the same spatial frequencies with respect to each other. 

\begin{figure}
\includegraphics[width=1.0\columnwidth, clip]{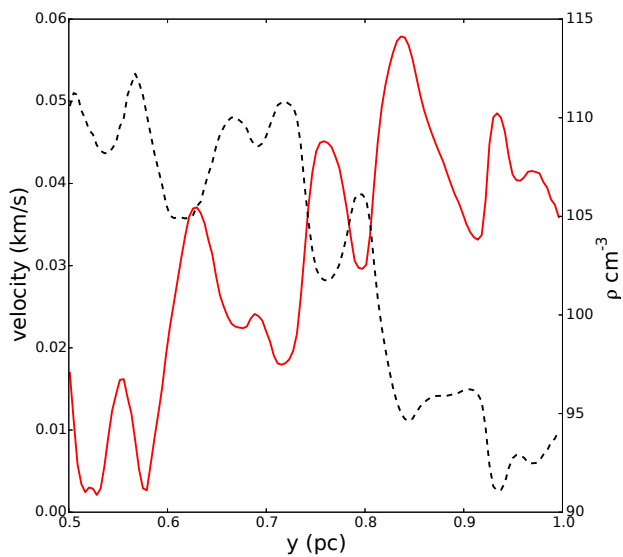}
\caption{Density (black dashed line) and velocity variation (red line) along one continuous structure from our 3D simulations with $\textit{z}$=0.25 pc in our fourth model. In agreement with Equation~\ref{bernoulli} an increase in velocity results in a decrease in density. These variations can be realized through the excitation of sausage MHD wave modes.}
\label{sausage}
\end{figure}

Secondary effects associated with the excitation of sausage MHD wave modes were also retrieved in our 3D simulations. In the data cube we first identified a continuous elongated structure and then examined the correlation between the velocity and density along that structure. In Figure~\ref{sausage} we plot the volume density (black dashed line) and the velocity (red line) throughout this striation. As expected from Bernoulli's principle the variation of density is in antiphase with the variation in velocity. A noticeable feature seen in Figure~\ref{sausage} is that a spectrum of sausage waves is excited, the superposition of which determines the final velocity and density variations. In this particular striation the effect is small compared to other regions inside the computational box. However, we chose to present results for this structure since it was one of the most clear cases and free of other effects, such as \Alf~waves, that could lead to a more complicated interpretation as to why these density and velocity variations occur.

Even for the example shown in Figure~\ref{sausage} where the change in velocity due to the excitation of sausage MHD waves is marginal, the effect may be observable. The velocity resolution of ALMA at a frequency of 110 GHz is 0.01 km/s. Thus, the velocity resolution from observations of $\rm{C^{18}O}$ ($\textit{J}$ = 1 $\rightarrow$ 0) emission at 109.78 GHz should be sufficient for the effect to be observed. Volume density variations could be derived by observing an additional $\rm{C^{18}O}$ transition and examining the line ratio.  	

The change in the cross sectional area of the striations is also of interest. If variations of the width along a single striation are found to be statistically significant, then the ratio of widths could be used to constrain the turbulent to ordered component of the magnetic field. The ratio of the width variance and the mean width should be proportional to the ratio of the turbulent and ordered component of the magnetic field:
\begin{equation}
\frac{A(r_{c})}{A(r_{c0})}=-\frac{B_c}{B_{c0}}
\end{equation}
(Grand et al. 2015) where $A(r_{c})$, $A(r_{c0})$, $B_c$ and $B_{c0}$ are defined as in Equations~\ref{cylindrical1} and~\ref{cylindrical2}. Variations of the width along a single striation could be measured with an algorithm such as FilTER (Panopoulou et al. 2016a) although, due to projection effects, this relation should be used with caution.  

\section{Summary and discussion}\label{discuss}

The current picture for the formation of striations includes streams that flow along magnetic field lines. We performed numerical simulations adopting two models involving such streamers, a model in which elongated structures are created as a result of a Kelvin-Helmholtz instability perpendicular to field lines and a new model in which striations are formed from the excitation of magnetosonic waves. We assessed the validity of each of our models by comparison between simulated and observational results based on four criteria:\begin{inparaenum}[\itshape a\upshape)] \item the contrast between minima and maxima in density and column density maps \item the spatial power spectrum in each velocity slice and in column density \item the kinematic properties (i.e. velocity range) \item whether the abundance of $\rm{CO}$ follows the total density~\end{inparaenum} and the contrast in abundance is significant for striations to be observed. We proved that flows, either sub-\Alf ic or super-\Alf ic, cannot reproduce the observed contrast even for huge velocity differences between ambient streamlines. The maximum possible contrast in the simulations involving flows is $\sim$ 0.03 \%. The mean contrast in observations is $\sim$ 25 \%.  

In our second model in which super-\Alf ic streamers flow along magnetic field lines, the contrast is low in both the 2D and 3D models. That is because the thickness of the LOS dimension is constant and thus cancelled out when computing the contrast in column density maps. As a result, projection effects are of minor importance and the observed contrast is an intrinsic density contrast rather than a geometrical effect.

Flows perpendicular to field lines is a mechanism that can also qualitatively produce elongated structures. However, due to development of turbulence, these structures are not long-lived. Furthermore, specific projection angles are required so that these structures are seen parallel to the magnetic field. A scenario in which this mechanism can simultaneously produce low density striations parallel to field lines and dense filaments perpendicular to the magnetic field is also difficult to realize. Finally and most importantly, such flows fail to reproduce the observed contrast of striations by more than three orders of magnitude. Overall, this mechanism cannot account for the formation of striations.

In the low column density parts of molecular clouds there is good coupling between matter and magnetic field since the degree of ionization is large. Hence, in a paradigm where magnetic field lines act as flux tubes and striations are formed from flows along field lines there must also be regions of stronger and weaker magnetic field. By definition however, such a configuration is equivalent to a wave travelling perpendicular to the long axis of the striations. The quasi-periodicity seen in perpendicular cuts in observations also suggests a formation mechanism which includes superposition of waves.

In contrast to streamers, a model including coupling of MHD waves is physical and can naturally explain the formation of striations. Furthermore, for a certain set of parameters the contrast can be up to 7 \%. Besides the large number of combinations (length of LOS dimension, density and magnetic field values) that can be realized and could alter this value, it would certainly be enhanced due to chemical effects. Radiative transfer effects might also be important in an intensity map. Therefore, we conclude that this model can account for the observed contrast. 

However, even in the 3D simulations performed here the total velocity range over which striations appear is a factor of $\sim$3 smaller compared to observations. There is a number of possibilities to explain this shortcoming. First, intrinsic magnetic field and density values in Taurus could be outside the parameter space considered. From the right panel of Figure~\ref{parameter_space} it can be seen that the lower the plasma $\rm{\beta}$ the larger the velocity range. The amplitude of the perturbation is also a key parameter that affects both the maximum range and the evolution of each velocity component. A second possibility is the existence of multiple sheet-like resonant structures along the LOS which move with respect to each other. In such a picture all of these sheet-like structures should have approximately the same boundaries in order for the same dominant frequency to be present in the spatial power spectrum in all velocity slices. We have demonstrated that when we consider a spectrum of \Alf~waves initially present in the system results are even more realistic and the velocity range increases by more than a factor of two compared to the simple case of having one \Alf~wave. Altering the distribution of power could further increase the velocity range. Although numerical dissipation does not significantly affect the velocity range (see Appendix~\ref{convergence}), the growth of transversal gradients could be affected by the boundary conditions. 


In a recent paper Hacar et al. (2016) presented a thorough analysis of the $\rm{CO}$ data also used in this paper. They concluded that suprathermal $\rm{CO}$ linewidths could be explained from optical-depth effects and multiple narrow
components the superposition of which act as a broadening mechanism. Based on this interpretation of $\rm{CO}$ linewidths they suggested that intrinsic gas motions were transonic. As a result, the velocities found for the majority of the simulations in the parameter study could be within observational ranges.

Despite the observational features for which our fourth model can account for, the question arises why we considered incompressible \Alf~waves initially present in the system rather than directly setting up compressible magnetosonic waves. \Alf~waves are exact solutions of the equations of ideal MHD and are thus longlived. Zweibel \& Josafatsson (1983) studied the damping mechanisms of MHD waves and naturally found that \Alf~waves were the longest lived mode. \Alf~waves can also act as the energy carriers from remote regions than the ones where striations are formed and are ultimately observed. Hence, in this model, they arise naturally as the source from which magnetosonic waves pump energy. The spectrum of \Alf~waves passing through an inhomogeneous region is of great importance. The energy distribution in the power spectrum will ultimately be a function of the properties of \Alf~waves initially present in the system. Consequently, the power spectrum of striations could be used to study the spectrum of \Alf~waves present in that region. We intend to return to the problem in follow-up publication with more 3D simulations and a larger parameter space. The effect that different dimensions and projection angles have on the power spectrum is also left for future study. 

Mouschovias (1987) predicted that torsional \Alf~waves can naturally be generated by the rotation of a clump and can also be trapped between magnetically linked clumps. Just as linear \Alf~waves, in the non-linear regime, torsional \Alf~waves can also excite fast magnetosonic waves (Tirry \& Berghmans 1997). As a result, striations connected to denser filaments could also be explained through the same mechanism. Thus, the interplay between \Alf~and magnetosonic waves along with acoustic waves and gravitational contraction along magnetic field lines is a promising scenario for explaining the overall gas-magnetic field morphology. Additional 3D simulation including gravity, will determine if the phase mixing between torsional \Alf~waves and fast magnetosonic waves can reproduce the observed properties of striations associated with denser parts of molecular clouds.

Elongated structures, usually referred to as fibers, have also been observed at high Galactic latitudes in the diffuse interstellar medium (see Clark et al. 2015 and references therein). Similar to striations, the magnetic field in these regions is well ordered and parallel to fibers which again exhibit quasi-periodicity. Thus, it is possible that striations and fibers share a common formation mechanism. 

\section*{Acknowledgements}

We thank G.V. Panopoulou, T. Mouschovias, V. Pavlidou and P. Goldsmith for useful comments and discussions. We also thank the anonymous referee for helpful suggestions that improved this paper. The software used in this work was in part developed by the DOE NNSA-ASC OASCR Flash Center at the University of Chicago. For post processing our results we partly used yt analysis toolkit (Turk et al. 2011). K.T. and A.T. acknowledge support by FP7 through Marie Curie Career Integration Grant PCIG- GA-2011-293531 ``SFOnset". A.T. and K.T. would like to acknowledge partial support from the EU FP7 Grant PIRSES-GA-2012-31578 ``EuroCal". Usage of the Metropolis HPC Facility at the CCQCN of the University of Crete, supported  by  the European Union Seventh Framework Programme (FP7-REGPOT-2012-2013-1) under grant agreement no. 316165, is also acknowledged.

\appendix
\section{Convergence tests}\label{convergence}

\begin{figure}\label{convergence_contrast}
\includegraphics[width=1.0\columnwidth, clip]{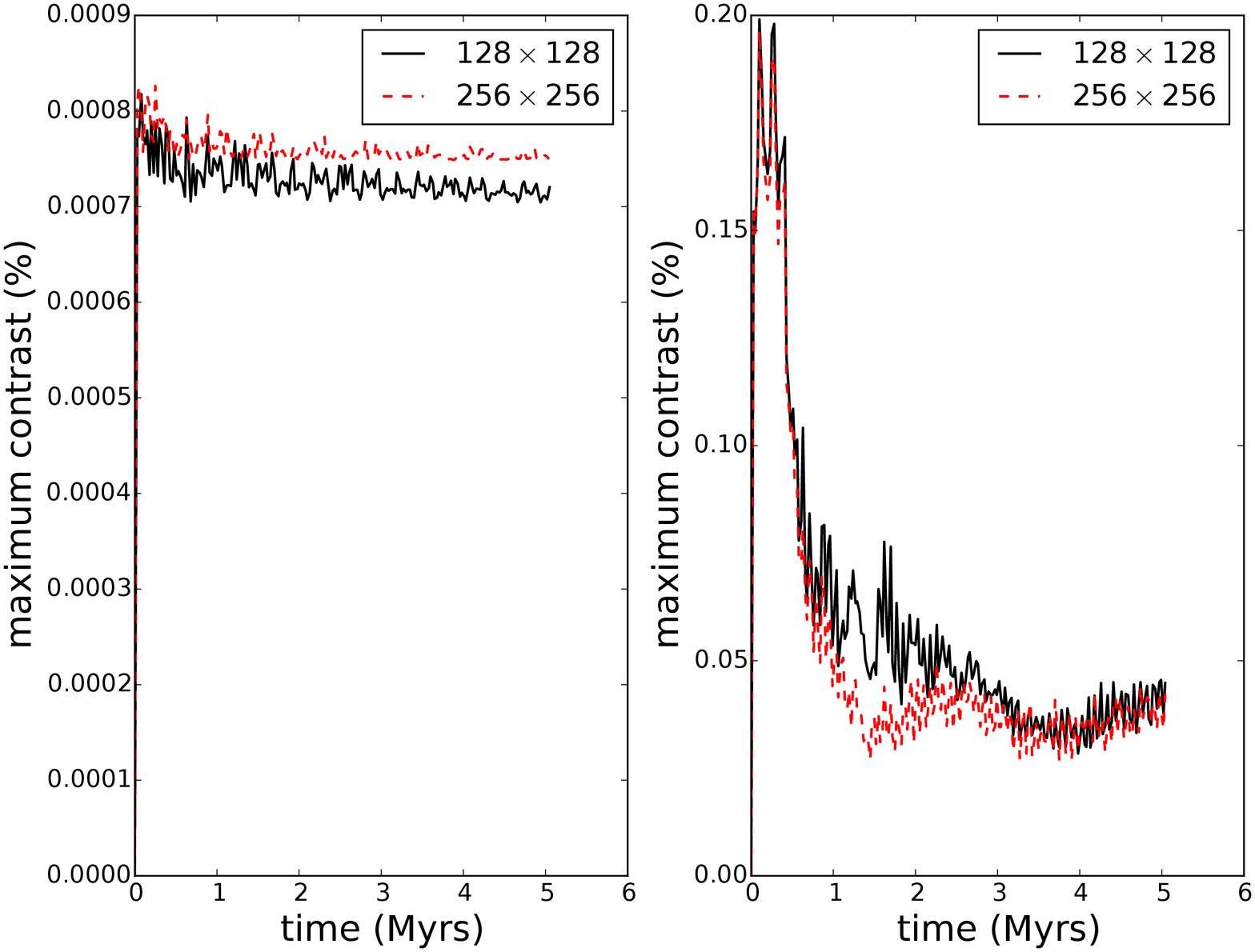}
\caption{Evolution of the maximum contrast as a function of time for the two streamer models. With the solid black line we plot our result from the simulation with $128\times128$ grid points and with the dashed red line the results from the simulation with a resolution of $256\times256$ points. }
\end{figure}

In Figure~\ref{convergence_contrast}1 we show the contrast as a function of time for both models involving flows along magnetic field lines for two different resolutions. In the left panel we plot our results for the sub-\Alf ic streamers and in the right results for the super-\Alf ic flow along field lines model. Because of the random number generator we used to initialize both problems different velocity values were assigned to grid cells which in turn led to the minor differences seen in both panels. However, in both cases, the lines follow the same trend and our results converge. The contrast has not be computed as described in \S~\ref{observations}. Instead, it is the maximum contrast inside the entire computation region. 

\begin{figure}\label{convergence_velocity}
\includegraphics[width=1.0\columnwidth, clip]{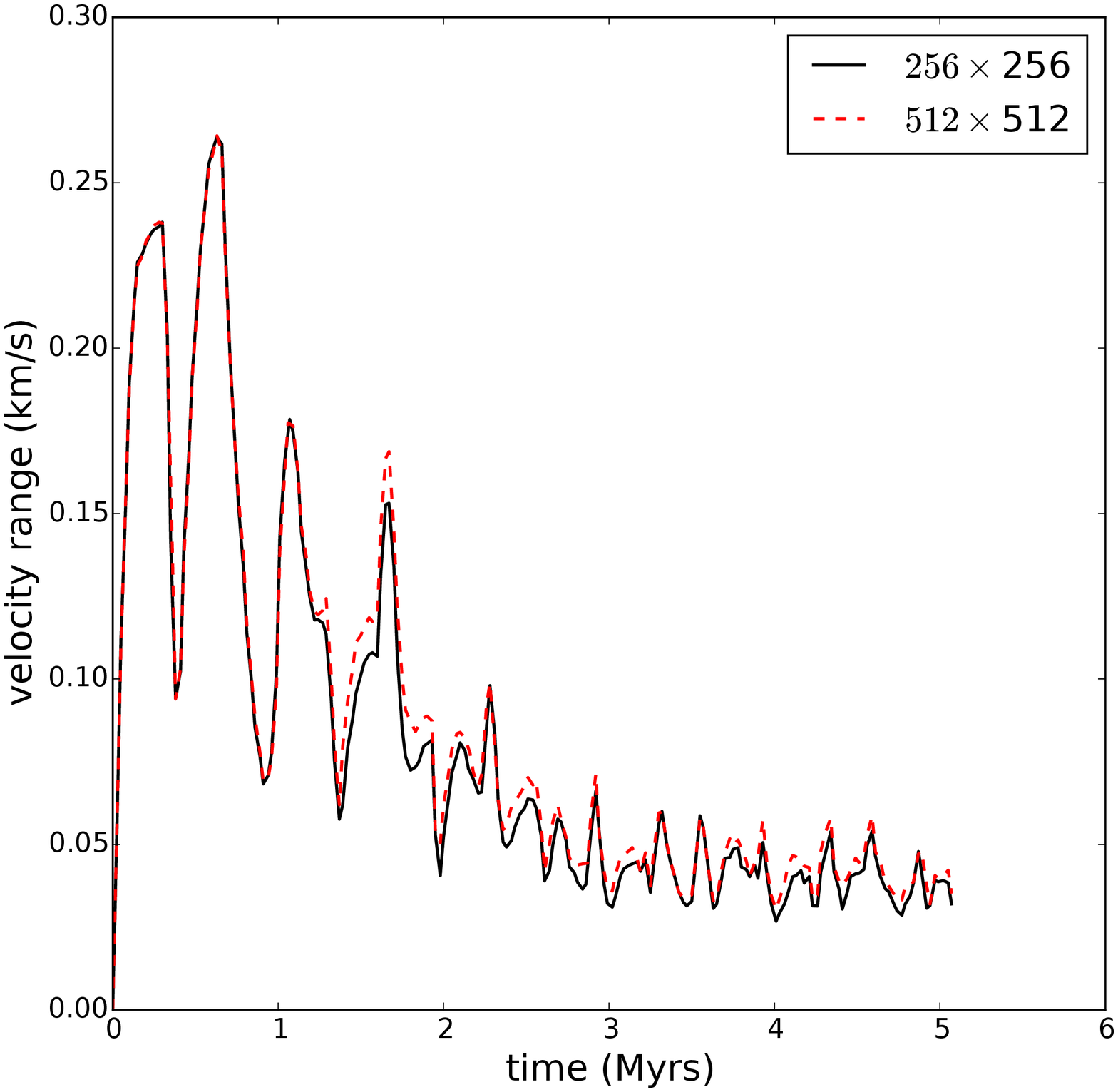}
\caption{Evolution of the \textit{x} velocity component as a function of time. With the solid black line we plot our result from the simulation with $256\times256$ grid points and with the dashed red line the results from the simulation with a resolution of $512\times512$ points. }
\end{figure}

In principle, numerical dissipation can stop the growth of transversal gradients very fast. If so, magnetosonic waves do not have time to pump enough energy from \Alf~waves for the observed velocities to be reached. However, as can be seen in Figure~\ref{convergence_velocity}2 where we show the evolution of the maximum velocity in the $\textit{x}$ direction for our fourth model and for two different resolutions, results converge. Therefore, numerical diffusivity does not significantly affect the velocity range. Even so, the growth of transversal gradients may still be affected by the boundary conditions.

\end{document}